\documentclass{article}

\usepackage{PRIMEarxiv}

\usepackage[utf8]{inputenc} 
\usepackage[T1]{fontenc}    
\usepackage{hyperref}       
\usepackage{url}            
\usepackage{booktabs}       
\usepackage{amsfonts}       
\usepackage{nicefrac}       
\usepackage{microtype}      
\usepackage{lipsum}
\usepackage{caption}
\usepackage{subcaption}
\usepackage[table,xcdraw]{xcolor}

\usepackage{wrapfig}
\usepackage{fancyhdr}       
\usepackage{graphicx}       
\usepackage{float}
\usepackage{amsmath}

\graphicspath{{figures/}}     

\usepackage{placeins}
\usepackage{soul,color}
\usepackage{multirow}

\title{Materials Transformers Language Models for Generative Materials Design: a Benchmark Study
\thanks{\textit{\underline{Citation}}: 
\textbf{N.F...J.H. . Materials Transformers: Language Models for Generative Materials Design: a Benchmark Study. DOI:000000/11111.}} 
}

\author{
Nihang Fu, Lai Wei, Yuqi Song, Qinyang Li \\
 Department of Computer Science and Engineering\\
  University of South Carolina\\
  Columbia, SC 29201
   \And
 Rui Xin, Sadman Sadeed Omee, Rongzhi Dong\\
 Department of Computer Science and Engineering\\
  University of South Carolina\\
  Columbia, SC 29201 \\  
   \And
  Edirisuriya M. Dilanga Siriwardane\\
 Department of Physics\\
  University of Colombo\\
 Colombo 03, Sri Lanka \\  
   \And
 Jianjun Hu *\\
 Department of Computer Science and Engineering\\
  University of South Carolina\\
  Columbia, SC 29201 \\
  \texttt{jianjunh@cse.sc.edu} \\
}

\begin{document}
\maketitle

\begin{abstract}
Pre-trained transformer language models on large unlabeled corpus have produced state-of-the-art results in natural language processing, organic molecule design, and protein sequence generation. However, no such models have been applied to learn the composition patterns of inorganic materials. Here we train a series of seven modern transformer language models (GPT, GPT-2, GPT-Neo, GPT-J, BLMM, BART, and RoBERTa) using the expanded formulas from material deposited in the ICSD, OQMD, and Materials Projects databases. Six different datasets with/out non-charge-neutral or balanced electronegativity samples are used to benchmark the performances and uncover the generation biases of modern transformer models for the generative design of materials compositions. Our extensive experiments showed that the causal language models based materials transformers can generate chemically valid materials compositions with as high as 97.54\% to be charge neutral and 91.40\% to be electronegativity balanced, which has more than 6 times higher enrichment compared to a baseline pseudo-random sampling algorithm. These models also demonstrate high novelty and their potential in new materials discovery has been proved by their capability to recover the leave-out materials. We also find that the properties of the generated samples can be tailored by training the models with selected training sets such as high-bandgap materials. Our experiments also showed that different models each have their own preference in terms of the properties of the generated samples and their running time complexity varies a lot. We have applied our materials transformer models to discover a set of new materials as validated using DFT calculations. 

\end{abstract}


\section{Introduction}

Out of the almost infinite chemical design space of inorganic materials, there are only 262,242 experimentally synthesized crystal structures as deposited in the ICSD database \cite{zagorac2019recent} for now in April 2022. Intelligent computational algorithms are strongly needed to navigate the huge uncharted chemical space for discovering novel materials. Currently, there are three major new materials discovery strategies: the first one is experimental tinkering in which researchers manipulate a given composition, synthesize it, and characterize its structure or function \cite{zunger2021understanding}; the second approach uses computational models to generate new compositions, and uses crystal structure prediction algorithms to predict their structures, and then uses DFT calculations to characterize their properties \cite{dan2020generative}; the third approach directly trains generative models for creating crystal structures for the downstream property prediction or simulation \cite{zhao2021high}. The first approach is too costly to explore the huge design space while the third approach is currently limited by the capability for existing crystal structure generation algorithms to generate stable structures. Considering the fact that most existing materials can be assigned to a limited number of prototypes, the emergence of template-based crystal structure prediction algorithms \cite{wei2021tcsp} has made it promising to explore new materials discovery using the composition generation and template-based crystal structure prediction (TCSP). 

Here we propose to use the deep learning language models for the generative materials composition discovery. Our work is inspired by the fact that a material composition or formula can be conveniently converted into a unique sequence of elements by assuming a specific element order (e.g. SrTiO3 --> Sr Ti O O O by the ascending element electronegativity) and that pretrained language models have been widely used in the generation of natural language texts, molecules, and protein sequences. These pretrained self-supervised learning models such as BERT \cite{devlin2018bert} and GPT-3 \cite{brown2020language} are able to learn language/chemical grammars \cite{wei2021frequency} for the text/molecule/protein generation \cite{rothe2020leveraging,li2021pretrained}. However, no such language models have been used for the generation of inorganic materials.  

There are several categories of pretrained language models for the text generation as reviewed in \cite{li2022learning} including masked language models, causal language models, prefix language models, and encoder-decoder language models. The masked language models such as BERT are trained by predicting the masked tokens using the contextualized information, which is not directly in alignment with the text generation task. They however can be used in the encoder and decoder part for text generation models by exploiting their excellent bidirectional encoding capacities. Causal language models such as GPT \cite{radford2018improving}, GPT-2 \cite{radford2019language}, and GPT-3 are trained to calculate/predict the probability of the occurrence of several words given all preceding words, making them ideal for the text generation. However, they have the weakness of neglecting the bidirectional information. Prefix language models such as UniLM \cite{dong2019unified} and XLNet \cite{yang2019xlnet} aim to combine the advantages of the bidirectional masked language models (LMs) and the unidirectional causal LMs in the text generation. A majority class of text generators such as T5 \cite{raffel2019exploring} and BART \cite{lewis2019bart} belongs to the encoder-decoder language models, which consist of stacks of both encoder and decoder layers. These models have all been used in the molecule or protein sequence generation but their performance for the inorganic composition generation is unknown. 

Despite that crystal inorganic materials, organic molecules, and proteins are all composed of atoms, they have a distinct difference in terms of their building blocks and topology: organic molecules consist of atoms connected by bonds while protein sequences are composed of chains of amino acids. In contrast, crystal materials are periodic structures, of which each unit cell contains a repeating structural pattern. There is no strict chain of proteins or connected components of organic molecules. On the other hand, they can all be represented as sequences, and the models for the generative design of proteins and molecules can serve as the source of reference of developing generative models for the material composition generation.

Deep language models have been used for the generation of molecules \cite{bagal2021molgpt,rothchild2021c5t5}. Inspired by the GPT model, Bagal et al. \cite{bagal2021molgpt} trained a transformer-decoder on the task that predicts the next token using masked self-attention for the generation of druglike molecules. Rothchild \cite{rothchild2021c5t5} proposed their novel self-supervised pretraining method that enables transformers to make zero-shot select-and-replace edits, altering organic substances toward the desired property, which shows better performance and potential than graph-based methods. In \cite{kim2021generative}, Kim et al. combined a transformer encoder with a conditional variational autoencoder (cVAE) to achieve the high performance molecule generation. A similar generative VAE was also proposed in \cite{dollar2021attention}. However, all these generative language models do not explicitly model the generative process and work more like black-box generators. In addition to VAE-based models, researchers also use some GAN-based or RNN-based models to generate molecules. Guimaraes et al. \cite{guimaraes2017objective} proposed a method that combines GANs and reinforcement learning to achieve that while reinforcement learning (RL) biases the data generation process towards arbitrary metrics, the GAN component of the reward function ensures that the model still remembers information learned from data. De Cao et al. \cite{de2018molgan} adapted GANs to operate directly on graph-structured data and design experiments on the QM9 chemical database to validate the performance of their model.

Language models have also been applied for the protein sequence generation \cite{madani2020progen,wu2020signal}. Madani et al. proposed an autoregressive transformer model named ProGen \cite{madani2020progen}, an 1.2 billion parameter conditional language model trained on a dataset of 280 million protein sequences. They also incorporated conditioning tags for taxonomic, functional, and locational information to enable the generation for targeted properties. Hesslow et al. \cite{hesslow2022rita} trained decoder-only transformer models without any conditioning information for the protein sequence generation. Ingraham \cite{ingraham2019generative} proposed a conditional generative model for the protein sequence generation given 3D structures based on graph representations. More recently, Ram and Bepler developed the MSA-to-protein transformer, a generative model of protein sequences conditioned on protein families represented by multiple sequence alignments (MSAs). Compared with previous generative studies usually without rigorous performance evaluations, Ferruz et al. \cite{ferruz2022deep} developed GPT-X based transformer models ProtGPT2 for generating de novo protein sequences. They found their generated protein sequences share some similarities with natural ones such as amino acid propensities and disorders. Linder et al. \cite{linder2020generative} developed a language model that can maximize the fitness and diversity of synthetic DNA and protein sequences. Overall, except for the normalizing flow models, most generative models, such as autoregressive models (RNN/LSTM/Transformers), VAE, and GAN, have been applied for the protein generation \cite{osadchy2021deep}. However, compared to molecule deep generation model studies, there lacks standard benchmark datasets and performance evaluation metrics, which significantly hinder the development of generative protein sequence models.  

Despite the success of deep language models in the protein and molecule sequence generation, no studies have been reported successfully applied deep language models to the inorganic materials composition generation except for our recent work on generative transformers \cite{wei2022crystal}. Here, we develop and evaluate six materials transformers based on different language models for the materials composition generation. We train all the models using the materials composition/formula data in the form of unlabeled expanded element symbol sequences from selected samples from ICSD/OQMD/Materials Projects (MP) databases. Compared to natural language texts, inorganic materials composition sequences have strong constraints among the elements due to the requirements to form chemically valid and structurally stable periodic structures. This involves complex atomic interactions from ionic or covalent bonds and/or oxidation states of constituent elements. Effective materials composition generation models have to learn complex local and long-range dependencies and generation contexts, for which transformer neural network models excel at detecting and modeling. Our language model-based composition generators have an advantage over the heuristic or data mining element substitution models \cite{hautier2011data,sun2019map} as they can consider the chemical context within the formulas rather than only element property compatibility. Our extensive generative composition design experiments show that the transformer-based materials generators can learn chemical grammars and achieve a high composition generation performance. Our additional experiments also show that materials transformers have different generation preferences or biases such as the tendency to generate materials compositions with >4 elements and with low numbers of atoms per element.

\section{Results}
\label{sec:headings}

\subsection{Pretrain Transformer language models for material composition generation }

We select 6 different transformer-based language models as implemented in the Huggingface package \cite{shen2020blank} to generate a series of Material Transformer (MT) generators. The group includes four GPT series language models, BART \cite{lewis2019bart}, and RoBERTa \cite{liu2019roberta}. In addition, we also use our previous work, BLMM \cite{wei2022crystal}, in our experiments to show the performance.

\begin{itemize}
    
    \item \textit{GPT (Generative Pre-trained Transformer) model} \cite{radford2018improving}: GPT is a transformer-based language model(LM) working on various NLP tasks with the unsupervised training. Original GPT uses 12-layer decoder-only transformer with masked self-attention, which is therefore powerful at predicting the next token in a sequence. Considering this property, we use GPT to generate our crystal formulas, and we call this GPT-based materials composition generation model as MT-GPT.
    
    \item \textit{GPT-2 model} \cite{radford2019language}: GPT-2 is a large transformer-based causal language model derived from GPT that is trained simply to predict the next word in large text corpses given all of the previous words within some text. GPT-2 models are trained with much more diverse text data with over an order of magnitude parameters than GPT. It uses a modified Byte Pair Encoding as input representation to combine the benefits of word-level LM with the generality of byte-level approaches. GPT-2 also has a few neural network architecture changes such as moving layer normalization to the input of each sub-block and adding a layer normalization to the final self-attention block. This model maps naturally to the crystal formula generation task. We call this GPT-2-based materials composition generation model as MT-GPT2. 
    
    \item \textit{GPT-J model} \cite{wang2021gpt}: GPT-J is an open-source version of the multi-head GPT-3 model \cite{brown2020language}. As an auto-regressive causal language model, it is originally used for the text generation task. GPT3 models use the same network architecture as GPT-2 except their use of alternating dense and locally banded sparse attention patterns in the transformer layers. 
    Since the core ability of GPT-J is to take a string of the text and predict the next token, GPT-J is good at generating texts from a prompt. In this paper, We call the GPT-J-based materials composition generation model as MT-GPTJ.
    
    \item \textit{GPT-NEO model}  \cite{gpt-neo,gao2020pile}: GPT-Neo is an implementation of GPT3-like causal language model using the Mesh-TensorFlow library. The original architecture of GPT-Neo is similar to GPT-3 except that GPT-Neo uses local attention in every other layers with a window size of 256 tokens. GPT-Neo is trained as an auto-regressive language model, which means that it can also predict the next token as previous models. We call this GPT-Neo-based materials composition generation model as MT-GPTNeo.
    
    \item \textit{RoBERTa model} \cite{liu2019roberta}: RoBERTa is a dynamic masking pretrained language model based on the BERT model \cite{devlin2018bert}. It achieves higher performance by applying an improved pretraining procedure over original BERT, which includes training with more epochs and larger mini-batches, removing the next sentence prediction objective, training on longer sequences and dynamically changing the masking pattern applied to the training data. We call this RoBERTa-based materials composition generation model as MT-RoBERTa.

    \item \textit{BART model} \cite{lewis2019bart}: BART combines the design of the transformer architecture with a bidirectional encoder (like BERT) and a left-to-right decoder (like GPT) to form a denoising autoencoder model. BART models are trained by corrupting text with a noising function and learning a model to reconstruct the original text. BART is particularly effective on the text generation task while still performing well on the comprehension tasks. We call this BART-based materials composition generation model as MT-BART.
    \item BLMM \cite{wei2022crystal} is a transformer-based generative language model for materials composition generation, which is based on the blank filling language model BLM \cite{shen2020blank}. It formulates the composition generation problem as a sequential probabilistic sequence rewriting problem, which allows it to directly model the generation process enabling its high interpretability and high efficiency in generation.

\end{itemize}

\subsection{De novo generative design of materials composition }

\paragraph{Model Training and hyper-parameters}

We prepare two sets of training datasets to train different MT models for the materials composition generation. The first set includes two datasets, the ICSD-mix dataset and the ICSD-pure dataset, which contain selected compositions from the ICSD database. The former includes samples that do not satisfy charge neutrality (CN) or balanced electronegativity (EB) while the latter contains only samples that satisfy both chemical criteria. To evaluate whether increasing the number of training samples can improve the generation performance, we also prepare the second set of datasets including Hybrid-mix, Hybrid-strict, and Hybrid-pure datasets with selected compositions from ICSD, Materials Projects (MP), and OQMD databases. The Hybrid-mix dataset includes all formulas from the three databases. The Hybrid-strict dataset are selected from the Hybrid-mix dataset, which contains only samples that satisfy charge neutrality and balanced electronegativity with the ICSD-oxidation assignments for all elements. The Hybrid-pure dataset has the same requirement except that the CN and EB are evaluated using more relaxed oxidation assignments for elements implemented as the default oxidation states in SMACT\cite{davies2019smact}. The detailed information of each dataset is in the Section \ref{subsec:dataset}, and the detailed sample numbers of each dataset are shown in Table\ref{tab:datasets}. 

For each dataset, we conduct hyper-parameter tuning to figure out the best hyper-parameters with reasonable tuning efforts (See Section \ref{subsec:para-tuning}). We determine the training epochs based on the check of the learning curves of the training and validation loss to avoid overfitting. For each trained transformer model, we generate 50,000 candidate compositions that contain more than one and less than nine elements and the total number of atoms is smaller than or equal to thirty. We then check the percentages of these samples that satisfy CN and/or EB.

\paragraph{Generation of hypothetical material compositions:}

\begin{figure}[ht!] 
    \begin{subfigure}[t]{0.33\textwidth}
        \centering
        \includegraphics[height=0.8\textwidth]{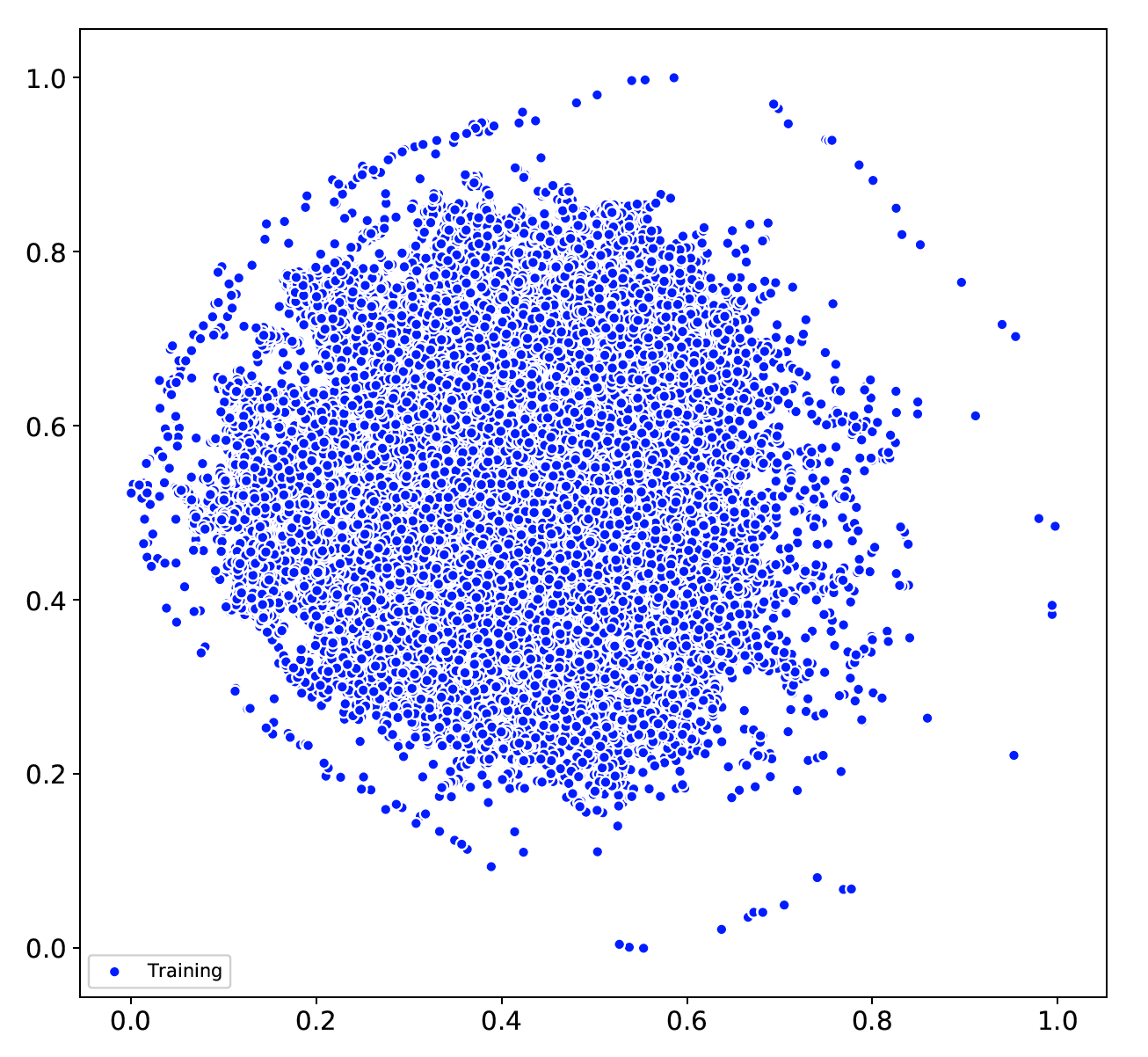}
        \caption{}
        \vspace{-3pt}
        \label{fig:train_tsne}
    \end{subfigure}
    \begin{subfigure}[t]{0.33\textwidth}
    \centering
        \includegraphics[height=0.8\textwidth]{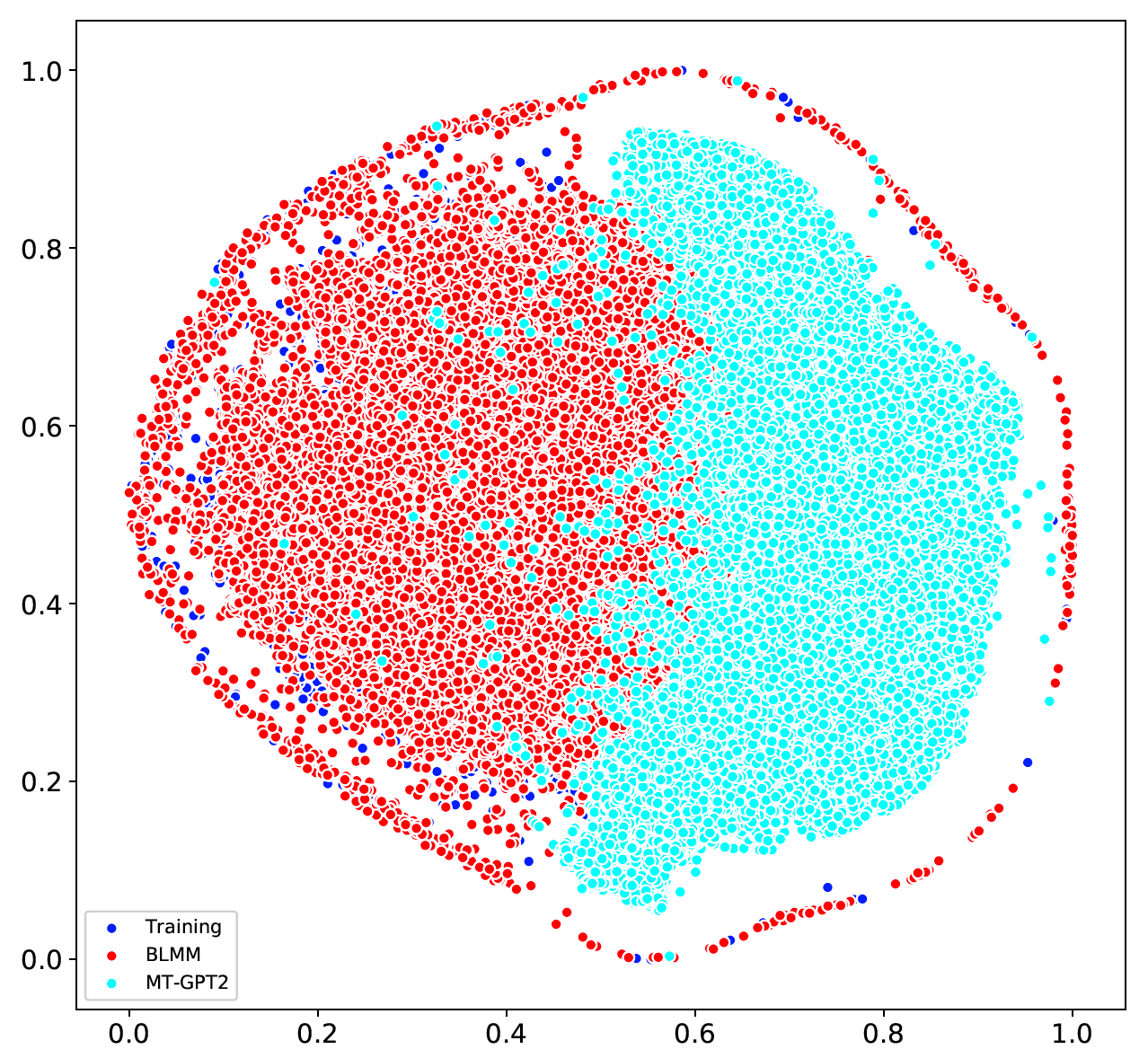}
        \caption{}
        \vspace{-3pt}
        \label{fig:mix_tsne}
    \end{subfigure}
    \begin{subfigure}[t]{0.34\textwidth}
         \raisebox{-0.085\height}{\includegraphics[height=0.81\textwidth]{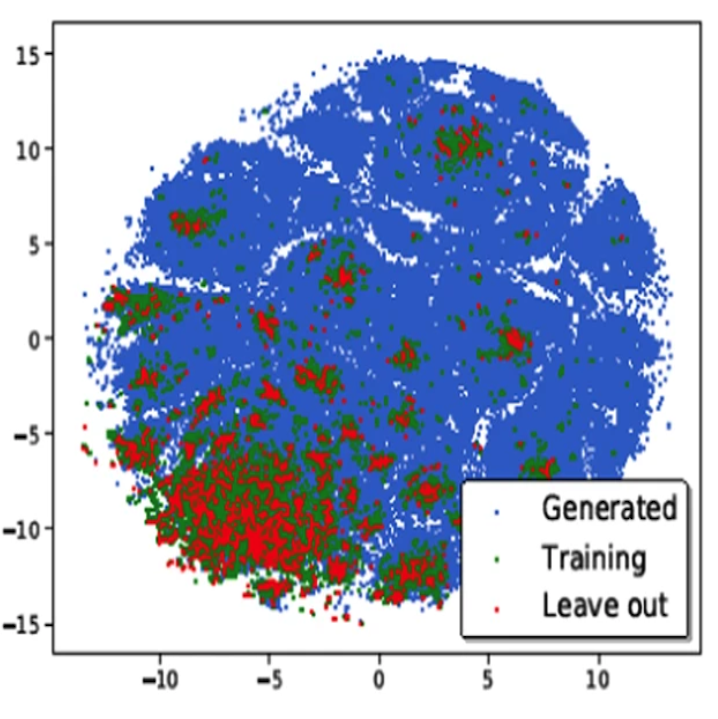}}
        \caption{}
        \vspace{-3pt}
        \label{fig:matgan_tsne}
    \end{subfigure}              

   \caption{The distributions of existing materials and hypothetical materials generated by BLMM, MT-GPT2, and MATGAN. The distributions are generated by calculating the one-hot representation for the compositions and then we use t-SNE to project them into 2-dimension space. (a) is the distribution of the training set. (b) is the distribution of the training set and generated samples by our BLMM and MT-GPT2. (c) is the distribution of the training set, the test set, and generated samples of MATGAN \cite{dan2020generative}. }
  \label{fig:distribution}
\end{figure}

To evaluate whether our language MT models can learn the chemistry of inorganic materials (compositions) and use it to generate valid hypothetical formulas, we first compare the distributions of the generated samples by BLMM, MT-GPT2, and MATGAN \cite{dan2019generative} with respect to the training set. We represent each formula using the one-hot encoding as described in \cite{dan2020generative} and then map all the sample matrix representations into two-dimension space using the t-SNE algorithm. The results are shown in Figure\ref{fig:distribution}. As shown in Figure\ref{fig:train_tsne}, we find that the compositions of the ICSD materials in the training set are not evenly distributed with a shift towards the left half space. Figure \ref{fig:mix_tsne} shows the distributions of the training set and samples generated by MT-GPT2 and BLMM, and the BLMM-generated samples have more overlaps with the training set compared to the MT-GPT2 generated ones. What is interesting is that both distributions of these two transformer generators are very different from the distribution of generated samples as distributed in Figure \ref{fig:matgan_tsne} in which the training samples are organized into several clusters corresponding to materials families and the known materials (training and testing samples) are only a tiny portion of whole composition space and the MATGAN tends to generate very different samples.

\begin{figure}[hbt!] 
\centering
        \includegraphics[width=0.9\linewidth]{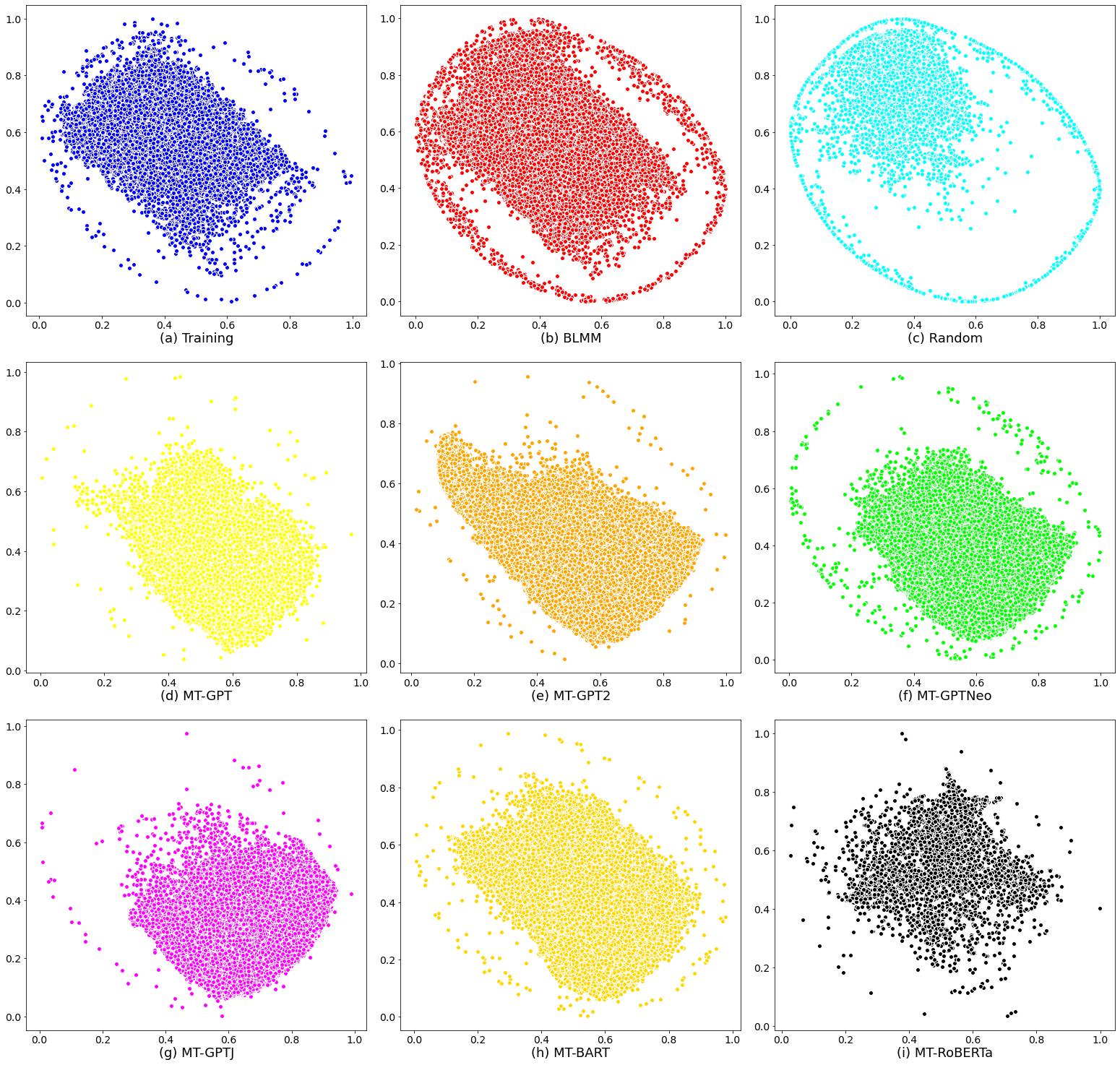}
        \vspace{3pt}
   \caption{The distributions of the training set and hypothetical materials generated by different materials transformers. The distributions are generated by calculating the one-hot representation for the compositions and then using t-SNE to project them into 2-dimension space. (a) Training samples; (b) BLMM; (c) Random algorithm; (d) MT-GPT; (e) MT-GPT2; (f) MT-GPTNeo; (g) MT-GPTJ; (h) MT-BART; (i) MT-RoBERTa. }
  \label{fig:dist_all}
\end{figure}

To further illustrate the composition pattern of the materials generators, we combine the training set and the generated samples by BLMM, pseudo-random algorithm, MT-GPT, MT-GPT2, MT-GPTNeo, MT-GPTJ, MT-BART, and MT-RoBERTa and conduct the t-SNE dimension reduction to map the 9 sets of samples into 2D space and then plot each set of compositions separately with the same coordinate system as shown in Figure \ref{fig:dist_all}. First, we find that the BLMM-generated samples in (b) show the highest similarity to the training samples in (a) compared to all other generators. The next most similar distribution with regard to the training samples is from MT-BART, which has a similar rectangular shape but with fewer samples in the upper-left area. The GPT series transformers tend to generate compositions with similar distributions with variation in several local areas in the chemical space. For example, they all have sparse samples in the upper-left area, which may be due to they have difficulty generating compositions with fewer than 5 elements (See Figure\ref{fig:dist_all}). The pseudo-random generator also shows a certain degree of distribution similarity with the training set because it uses its composition prototypes as the templates for composition generation, which means that it has the same number of binary, ternary, and quaternary samples as the training set. Finally, it is found that the samples generated by MT-RoBERTa have a very different distribution. This model is also the one that has the most difficulty to generate compositions with <=30 atoms with <=8 elements.

\subsection{Evaluations of materials transformer generation performance using validity, uniqueness, recovery rate, and novelty}

We evaluate the performance of our generative models for materials composition design and compare with the baseline random formula generator using four evaluation criteria including validity, uniqueness, recovery rate, and novelty as described in the Section \ref{subsec:criteria}. 

\begin{figure}[ht!] 
    \centering
        \includegraphics[width=0.8\textwidth]{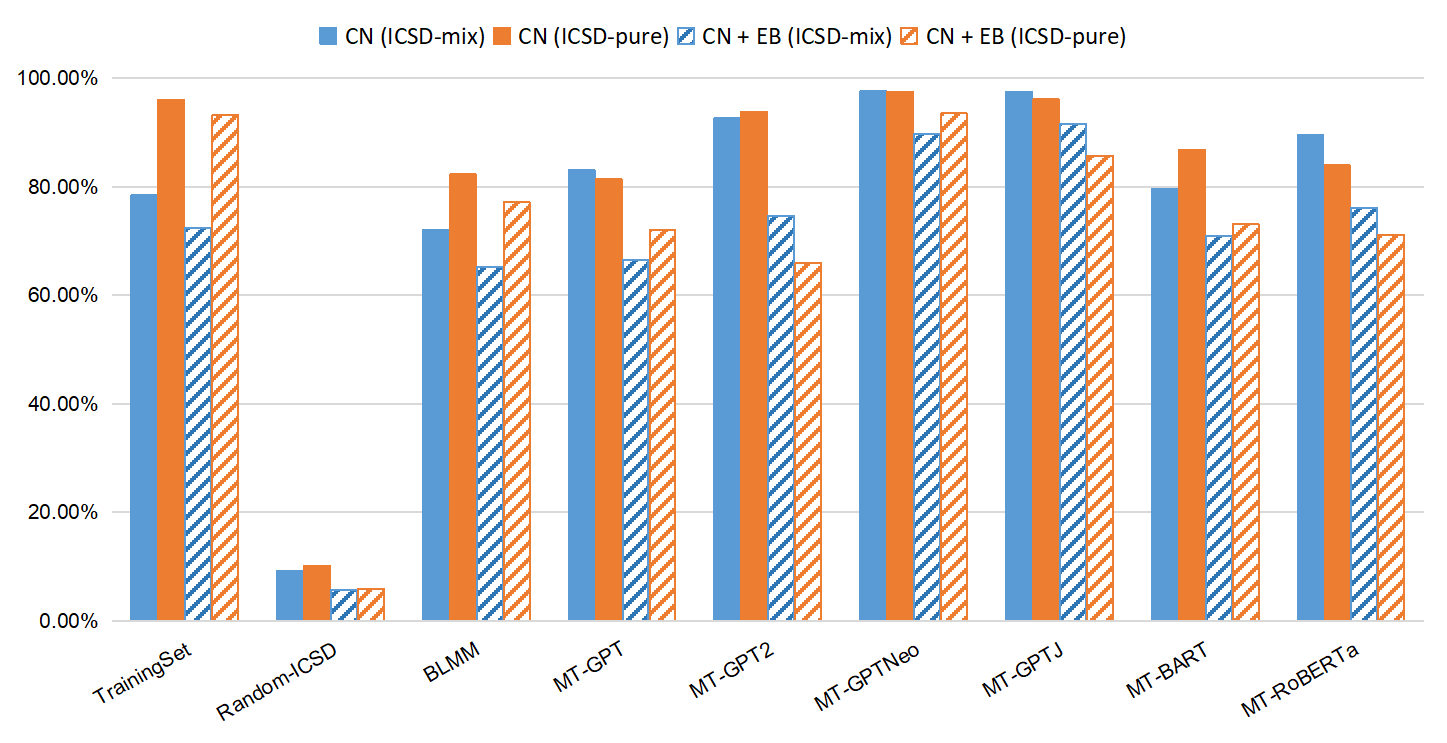}
   \caption{The comparison of the validity of materials composition generators on the ICSD-mix and ICSD-pure datasets. The percentages of CN and CN+EB samples out of the training set and all generated samples by the generator models are used to represent validity.}
  \label{fig:validity}
\end{figure}

\paragraph{The CN and EB performance}
Figure\ref{fig:validity} shows the composition generation validity performance of seven transformer-based models compared them with the pseudo-random generator as evaluated on the ICSD-pure (37459 samples with 95.94\% CN and 93.12\% CN+EB ) and ICSD-mix ( 50755 samples with 78.40\% CN and 72.36\% CN+EB) datasets. Note that all the percentages are calculated on the generated samples with less than or equal to 9 elements and 30 atoms. First, the CN and EB percentages of generated samples of all seven transformer models range from 65.87\% to 97.54\%, which are more than six times higher compared to the ones of the random generator (max CN: 10.13\% and max CN+EB: 5.91\%). Out of the seven models, we find that overall, the MT-GPTNeo and MT-GPTJ have the best validity performance, which is followed by MT-GPT2, MT-BART, and MT-RoBERTa. The BLMM and MT-GPT models tend to have the lowest validity, even though their gaps with MT-BART and MT-RoBERTa are small. However, a close investigation of the filtering process shows that more than 45\% of generated compositions have been filtered out for MT-BART, MT-RoBERTa, and the GPT series models while the BLMM filters out less than 0.3\% samples, as shown in Figure \ref{fig:elemenet_dist}, MT models tend to generate formulas more than 5 elements, but we would like to filter these long formulas in the filtering process. Especially, we would like to note here that MT-RoBERTa's performance evaluated here is calculated on less than 10,000 generated samples as it has difficulty generating 50,000 candidate samples with atom numbers less than or equal to 30 and fewer than 9 elements within a reasonable amount of time (most samples are filtered out). The non-BLMM models tend to generate compositions with more than 8 elements. In addition, these seven models show a big difference in terms of sample generation speed as shown in Figure \ref{fig:time}. 

Another performance trend we find is that the CN/EB validity percentages of samples generated by models trained on the ICSD-pure dataset are in general higher than those by the models trained on ICSD-mix except for the case of the CN+EB percentages of MT-GPT2, MT-GPTJ, and MT-RoBERTa. After close examination, we find these exceptions are mainly due to the filtering process as discussed above. 

We also compare the validity performance of our MT-GPTNeo model with our previously developed MATGAN models that are based on the generative adversarial network \cite{dan2020generative}. We find our MT-GPTNeo model achieves 97.54\% CN and 84.64\% EN (for the filtered samples with less than 31 atoms and 9 elements) compared to  80.3\% CN and 70.3\% EB achieved by the GAN model trained with ICSD-mix dataset, showing our models have about 14-17\% advantage. Similar performance advantages are also observed for our models trained with the ICSD-pure dataset. However, we want to note that this performance advantage is evaluated over the filtered samples. 

To check whether increasing the training samples can improve the generator performance, we train our models on the three Hybrid datasets including Hybrid-mix, Hybrid-pure, and Hybrid-strict which have 398033, 244281, and 202139 training samples respectively. The model hyper-parameters are also tuned with the new datasets. The final performances are shown in Table \ref{tab:hybrid-performance}. First, we find that the CN and EB percentage of the training set of Hybrid-mix is much lower than those of the ICSD-mix with only 61.67\% CN and 50.61\% respectively. However, the validity percentages of all models trained on the Hybrid-mix dataset are much higher than those of the training set. Especially GPT series models achieve CN percentages higher than 92\% and EB percentages higher than 78\%. These unexpected high validity percentages are due to the filtering process: they are good at generating chemically valid material compositions with less than 9 elements and 31 atoms while they also generate a large percentage of large formulas with many atoms (>30) and elements (>8). Then, the Hybrid-pure dataset has higher CN (83.48\%) and EB (75.94\%) than the Hybrid-mix dataset. And it helps to improve the CN percentages of MT-GPTNeo, MT-GPTJ, MT-BART, and MT-RoBERTa and the EB percentages of MT-GPTNeo and Mt-BART. Finally, the training set of the Hybrid-strict dataset has the highest CN (99.54\%) and EB (99.25\%). Almost all models except MT-GPTNeo achieve the highest CN and EB, indicating that high quality training samples contribute to the high validity performance of the trained models. Overall, we found that the MT-GPTJ has the best performance with CN of 96.68\% and EB of 90.33\% on Hybrid-mix, CN of 97.54\% on Hybrid-pure, and CN of 97.61\% on the Hybrid-strict dataset. We also found that MT-BART and MT-RoBERTa on the Hybrid-mix dataset have the lowest validity performance compared to those of GPT series.

As we mention earlier, the Hybrid datasets are datasets with samples ranging from 200,000 to 40,000, while the ICSD datasets are datasets with samples from 35,000 to 53,000. Therefore, we also compared the model performance trained by the Hybrid datasets and the ICSD datasets to investigate the effect of the quantity of samples on the generation performance. It is found that the CN and EB percentages of MT-GPT are not only increased from 81.01\% and 66.54\% to 92.24\% and 78.29\% on the Hybrid-mix, but also increased from 81.47\% and 71.96\% to 91.51\% and 77.76\% on Hybrid-pure. For MT-GPT2 and MT-RoBARTa, their CN and EB percentages have improved on the Hybrid-mix dataset compared to the ICSD-mix dataset. As for MT-GPTJ and MT-Bart, they achieve better CN and EB percentages on the Hybrid-pure dataset than those on the ICSD-pure dataset.

\begin{table}[]
\centering
\caption{Comparison of generator performances of models trained with Hybrid datasets.}
\label{tab:hybrid-performance}
\begin{tabular}{
>{\columncolor[HTML]{FFE599}}l |cc|cc|cc}
\hline
            & \multicolumn{2}{c|}{\cellcolor[HTML]{FFE599}Hybrid-mix}               & \multicolumn{2}{c|}{\cellcolor[HTML]{FFE599}Hybrid-pure}             & \multicolumn{2}{c}{\cellcolor[HTML]{FFE599}Hybrid-strict}            \\ \hline
Model       & \multicolumn{1}{c|}{CN}                     & CN + EB                & \multicolumn{1}{c|}{CN}                     & CN+EB                  & \multicolumn{1}{c|}{CN}                     & CN+EB                  \\ \hline
TrainingSet & \multicolumn{1}{c|}{61.67\%}                & 50.61\%                & \multicolumn{1}{c|}{83.48\%}                & 75.94\%                & \multicolumn{1}{c|}{99.54\%}                & 99.25\%                \\ \hline
MT-GPT      & \multicolumn{1}{c|}{92.24\%}                & 78.29\%                & \multicolumn{1}{c|}{91.51\%}                & 77.76\%                & \multicolumn{1}{c|}{92.21\%}                & 84.57\%                \\ \hline
MT-GPT2     & \multicolumn{1}{c|}{92.96\%}                & 79.79\%                & \multicolumn{1}{c|}{87.82\%}                & 70.83\%                & \multicolumn{1}{c|}{96.99\%}                & { \textbf{92.81\%}} \\ \hline
MT-GPTNeo   & \multicolumn{1}{c|}{93.84\%}                & 84.37\%                & \multicolumn{1}{c|}{97.29\%}                & { \textbf{91.40\%}} & \multicolumn{1}{c|}{94.69\%}                & 77.39\%                \\ \hline
MT-GPTJ     & \multicolumn{1}{c|}{{ \textbf{96.98\%}}} & { \textbf{90.33\%}} & \multicolumn{1}{c|}{{ \textbf{97.54\%}}} & 87.05\%                & \multicolumn{1}{c|}{{ \textbf{97.61\%}}} & 91.22\%                \\ \hline
MT-BART     & \multicolumn{1}{c|}{81.10\%}                & 62.83\%                & \multicolumn{1}{c|}{85.23\%}                & 70.07\%                & \multicolumn{1}{c|}{88.01\%}                & 80.47\%                \\ \hline
MT-RoBERTa  & \multicolumn{1}{c|}{71.16\%}                & 61.00\%                & \multicolumn{1}{c|}{84.66\%}                & 60.26\%                & \multicolumn{1}{c|}{93.68\%}                & 80.81\%                \\ \hline
\end{tabular}
\end{table}

\FloatBarrier

\begin{figure}[ht!] 
  \centering
  \includegraphics[width=\linewidth]{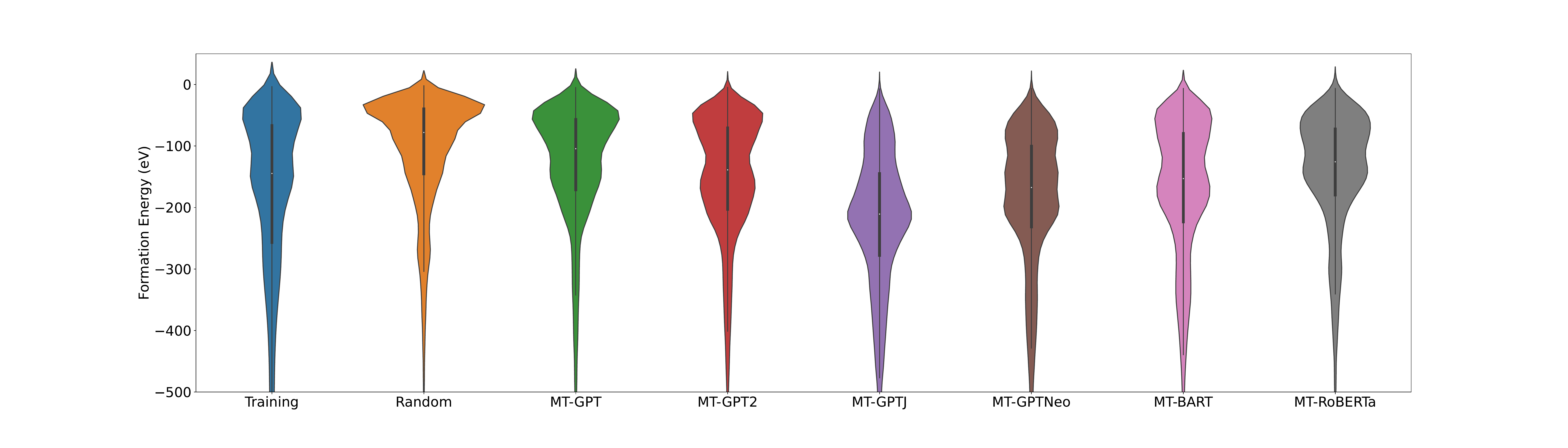}
   \caption{The comparison of formation energy distributions of the training samples, samples from the pseudo-random generator, and generated samples by different models trained on the ICSD-pure dataset.} 
  \label{fig:formenergy_dist}
\end{figure}

\paragraph{The stability performance} We also evaluate the materials composition generation validity performance by checking the stability of the generated compositions by all MT models and the pseudo-random generator as partially represented by their predicted  formation energies. We train a Roost \cite{goodall2020predicting} based formation energy machine learning prediction model (see Section \ref{subsec:formation_energy}) using all the MP samples. The formation energy distributions of the ICSD-pure training set, the generated samples of all MT models and random samples are shown in Figure\ref{fig:formenergy_dist}. First, we find that the formation energies of the training set are mostly less than zero eV and the shape of the distribution looks like a standard violin while the distribution of the random samples is very different from the training set with a large percentage of samples located in the near-zero eV area. Out of the seven MT models, the generated samples of MT-GPT and MT-RoBERTa are of lower quality as they are more located in the high-energy region. While the BLMM model's samples show higher similarity in terms of the distribution shape to the training samples, the samples generated by MT-GPT2, MT-GPTNeo, and MT-BART tend to have lower formation energy. The best quality of generated samples as represented by the predicted formation energy comes from the MT-GPTJ model, which has a peak density of samples around -200 eV. Overall, Figure \ref{fig:formenergy_dist} shows that our MT models can generate chemical valid samples with negative formation energies. 

\paragraph{The recovery rate performance} Another way to check the validity of generators is to check the number of generated samples that do not exist in the training set and exist as known materials in the leave-out dataset or third-party experimental or computational databases. The criterion is sometimes calculated as the recovery rate. We check the 56,162 generated samples of MT-GPTJ and find that 196 samples exist in one of the three databases, ICSD/OQMD/MP. In terms of holdout recovery rate, our model's performance (overall 0.81\%) is much lower compared to those of the BLMM model, which achieves 100\% for binary materials, 63.37\% for ternary materials, and 29.17\% for quaternary ones. It is also lower than the MATGAN model which has binary, ternary, and quaternary compounds recovery rates of only 82.7\%, 31.2\%, and 5.2\%. The reason is that our transformer-based language models tend to generate compositions with more than 5 elements (see Figure \ref{fig:atom_dist}).

\begin{figure}[ht!] 
  \centering
  \includegraphics[width=0.6\linewidth]{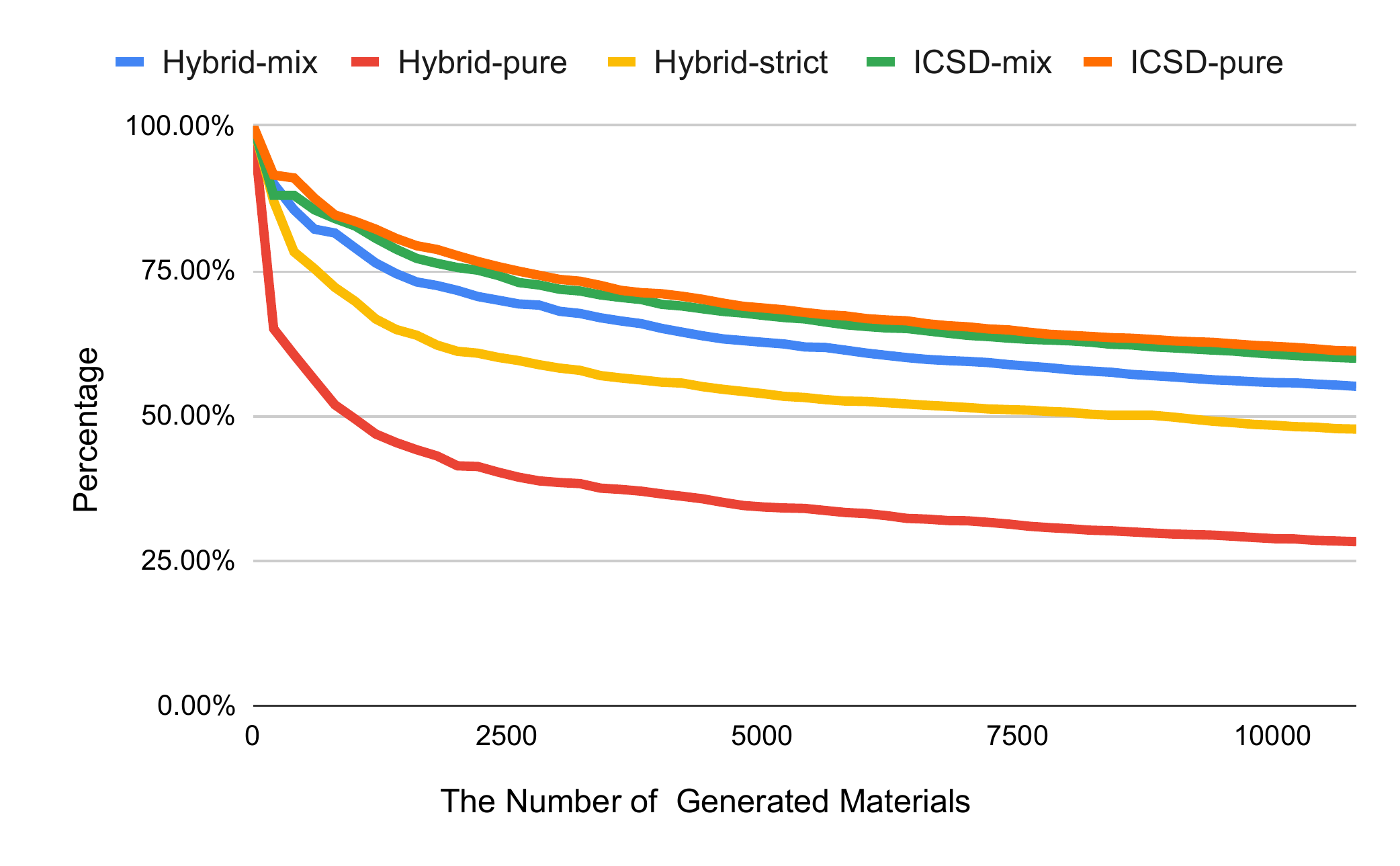}
\caption{The uniqueness of the MT-GPTNeo model trained on ICSD-pure dataset.}
  \label{fig:uniq}
\end{figure}

\paragraph{The uniqueness performance} Another important performance measure of generative models is the uniqueness, which represents the percentage of unique samples out of all generated samples \cite{polykovskiy2020molecular}. Here for five MT-GPTNeo models trained on the five datasets, we calculate the uniqueness percentages for every 200 generated samples up to 10,800 samples with <=30 atoms and <=8 elements. The results are shown in Figure\ref{fig:uniq}. First, we find that all five MT-GPTNeo models show high uniqueness: after generating 10,800 filtered samples, the uniqueness percentages remain around or above 50\%: ICSD-mix (59.91\%), ICSD-pure (61.38\%),  Hybrid-mix (55.06\%), Hybrid-strict (47.69\%), and Hybrid-pure (28.27\%). Another interesting observation is that the models trained on the ICSD-mix and ICSD-pure datasets dominate in terms of the uniqueness, which is probably due to the fact that these two datasets have much fewer training samples (50755 and 37459 versus more than 200,000 for hybrid datasets). The more training samples, the more strict language constraints the models learn, and thus the lower the diversity/uniqueness in the generated samples. Among all the three Hybrid models, the model trained with the Hybrid-mix has the highest uniqueness since it contains diverse samples that do not satisfy CN/BN chemical rules. The uniqueness difference of these three models can be attributed to their different distributions among the training sets. 

\paragraph{The novelty performance} We also check the novelty of the composition generators, which calculates the percentages of samples that do not exist in the training set. All our six models achieve more than 97\% overall novelty when trained over the ICSD-pure dataset. In contrast, the BLMM model achieves 97.66\%, 96.11\%, and 95.55\% novelty for binary, ternary and quaternary compounds respectively, indicating comparable or slightly better capability to generate new materials.

\FloatBarrier

\subsection{Conditional generative design of materials compositions with high bandgap}
\label{subsec:bandgap}

\begin{figure}[ht]
  \centering
  \includegraphics[width=0.7\linewidth]{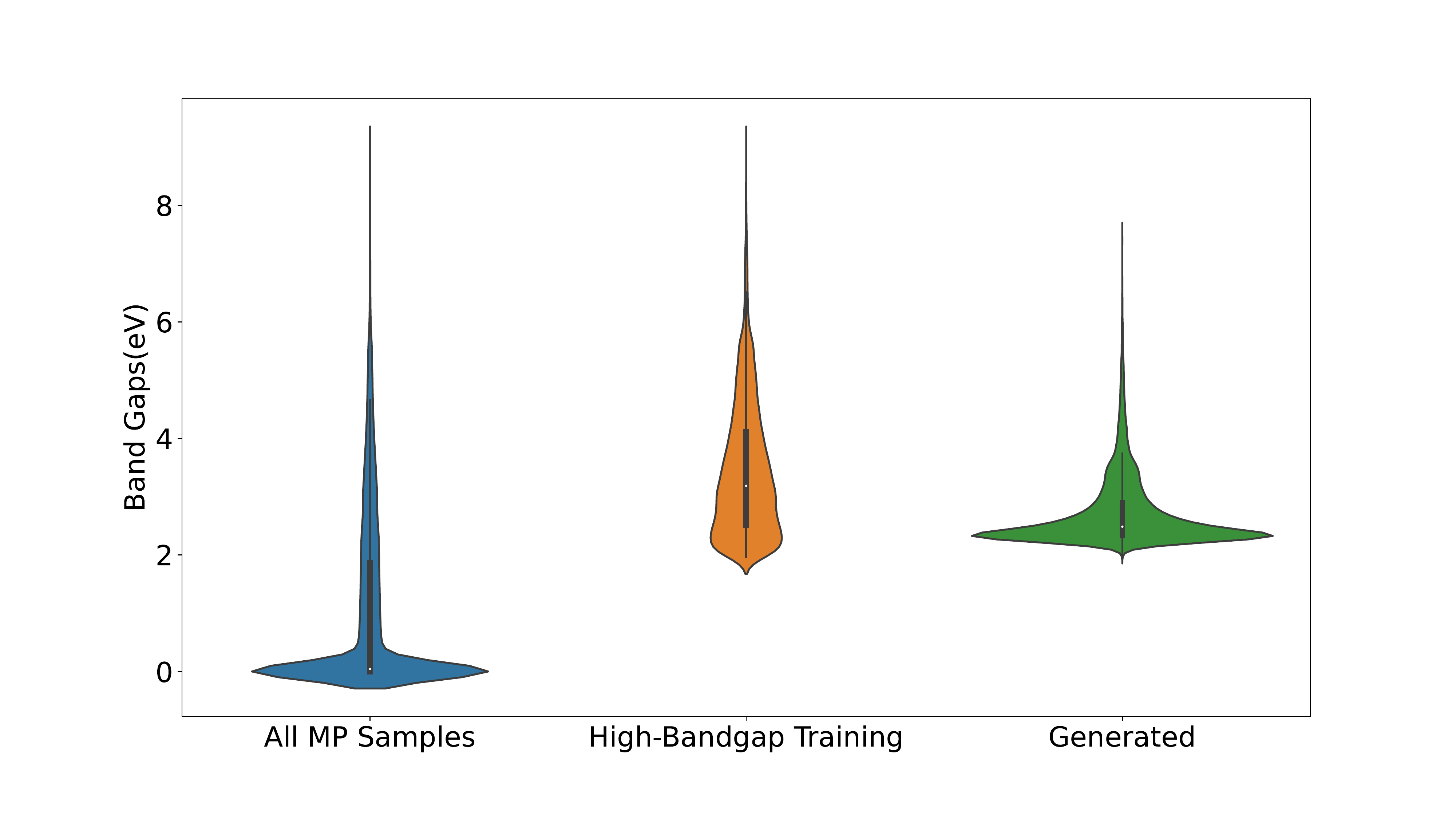}
  \vspace{-5mm}
  \caption{Band gap distribution for (1) the whole Materials Projects (MP) materials, (2) the training set of high-bandgap MP materials for the MT-GPT2 model; (3) the generated samples from the MT-GPT2 models trained on the MP dataset. The band gap distribution of the generated ones is much closer to the training set than the whole MP dataset, they tend to have higher band gaps.}
  \label{fig:bandgap}
\end{figure}

Conditional generation capability is highly desirable for function-oriented computational materials design. \cite{flam2022language} used this evaluation method for benchmarking molecule generator models. Therefore, to evaluate whether our language model can capture the composition rules for assembling high-bandgap materials to directed high-bandgap materials generation, we collect 30,000 formulas with band gaps above 1.98 eV from the Materials Projects database (for those formulas with multiple phases, we include it if it has one phase with band gap greater than 2.0 eV). We call this Bandgap-30K dataset. We train an MT-GPT2 composition generator and use it to generate 100,000 formulas from which 87,233 compositions satisfy the charge neutrality and balanced electronegativity requirements. We then use the composition-based band gap prediction model (See Section \ref{subsec:formation_energy}) to predict the band gaps of these filtered hypothetical material compositions and plot their distribution against the band gap distributions of the training set and the whole MP samples. As shown in Figure \ref{fig:bandgap}, the band gap distribution of our hypothetical materials is much closer to the high-bandgap training set compared to the band gap distribution of all MP samples, which indicates that the MT-GPT2 bandgap model has learned the implicit rules to generate high-bandgap materials. 
\vspace{-4mm}

\subsection{New materials predicted by MT-GPT2 and validated using DFT}

\begin{figure}[ht]
  \centering
  \includegraphics[width=0.6\linewidth]{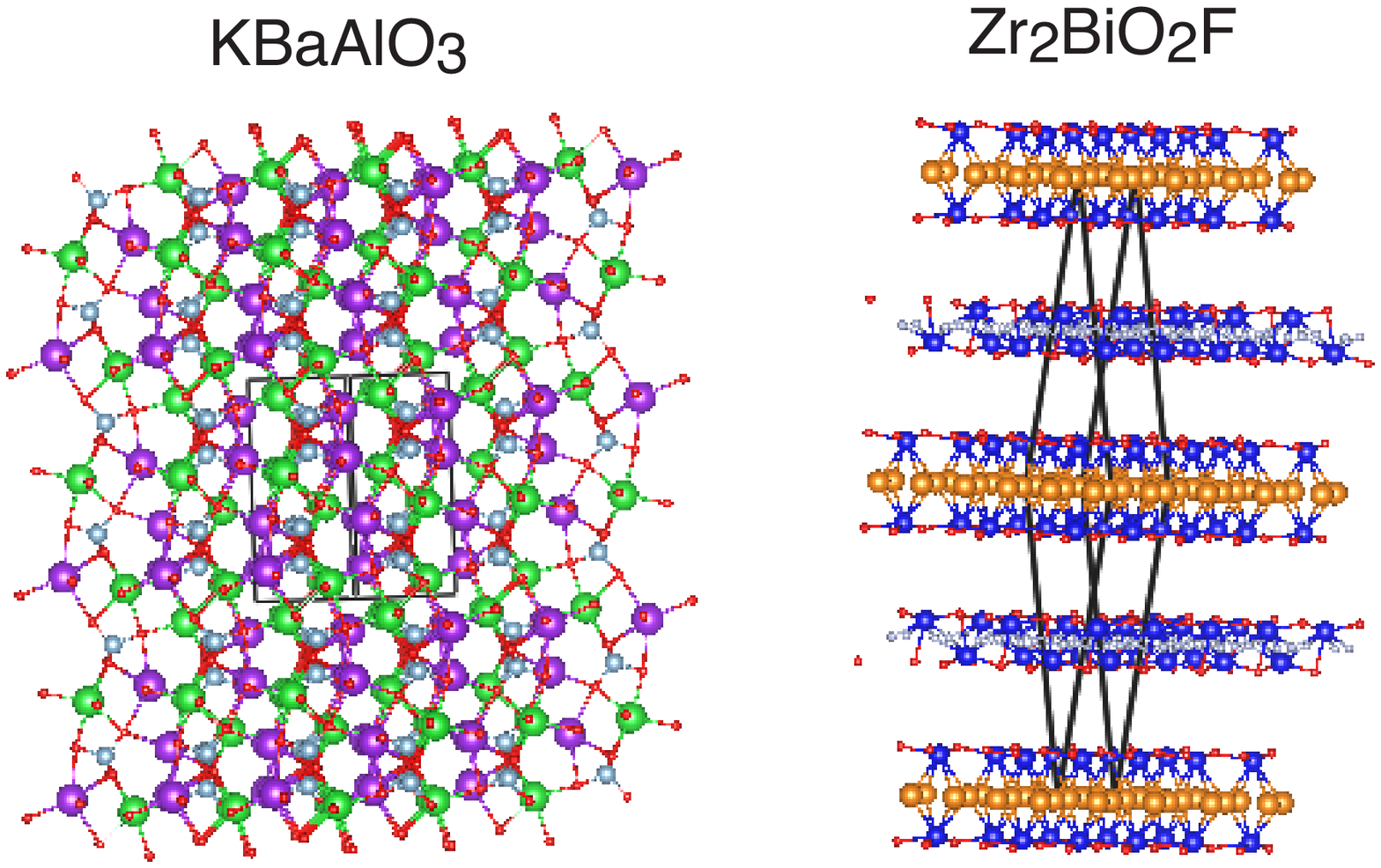}
  \caption{Candidate new structures discovered by our MT-GPT2 model with zero e-above-hull.}
  \label{fig:structuresfound}
\end{figure}

 We use the Bandgap-30K dataset to train a MT-GPT2 model and use it to generate more than 100,000 material compositions. Then we predict their formation energy using the composition based formation energy prediction model (See Section \ref{subsec:formation_energy}). When we get the formation energies, we calculate their total energy and predict their e-above-hull energies to rank these candidates. We then pick the top 100 formulas with the lowest predicted e-above-hull energy and apply our previous TCSP, a template based crystal structure prediction algorithm \cite{wei2021tcsp}, to obtain the structures. For the predicted structures with the best quality scores, we run DFT relaxation to get the final structures and to calculate their formation energy and e-above-hull energy (see Section \ref{subsec:DFT}). Table \ref{tab:finding} shows the top 20 discovered new materials along with their formation energies. Out of the predicted structures, we identify two new crystal materials with the e-above-hull energy of 0 eV, and their structures are shown in Figure \ref{fig:structuresfound}.

\begin{table}[]
\centering
\caption{Twenty materials found with negative formation energy ($E_\mathrm{form}$) using DFT}
\label{tab:finding}
\vspace{5pt}
\begin{tabular}{|l|l|l|l|}
\hline
Formula  & $E_\mathrm{form}$ & Formula   & $E_\mathrm{form}$ \\ \hline
SrAlClO2 & -3.0077           & BaSrTiO3  & -2.5268           \\ \hline
LiMgBrF2 & -2.9039           & LiScNiF3  & -2.5214           \\ \hline
BaScSeF2 & -2.8662           & ScBeOF    & -2.5074           \\ \hline
AlBrF2   & -2.8367           & KSrAlO3   & -2.5024           \\ \hline
LiMg2IF4 & -2.8118           & AlFeOF3   & -2.5015           \\ \hline
KSrScO3  & -2.7351           & Sc2AlZnO4 & -2.4719           \\ \hline
BaScO2   & -2.6801           & LiBeOF    & -2.4639           \\ \hline
KBaScO3  & -2.6221           & SrBeSeF2  & -2.4584           \\ \hline
KBaAlO3  & -2.5924           & MgVHF4    & -2.4521           \\ \hline
RbBeOF   & -2.5392           & KSrZrO3   & -2.4300           \\ \hline
\end{tabular}
\end{table}

\FloatBarrier

\section{Discussion}

We have applied a series of transformer-based pretrained language models for the generative design of inorganic materials compositions. Our extensive experiments over five datasets with diverse sample sizes and quality (in terms of charge neutrality and balanced electronegativity) show that these materials transformer models are competent in generating chemically valid compositions or formulas while most of them have their own generation preference or bias. We find that MT-GPTJ overall has the best generation performance (after simple filtering) in terms of validity and generation speed. 

\begin{figure}[ht]
  \centering
 \includegraphics[width=0.6\linewidth]{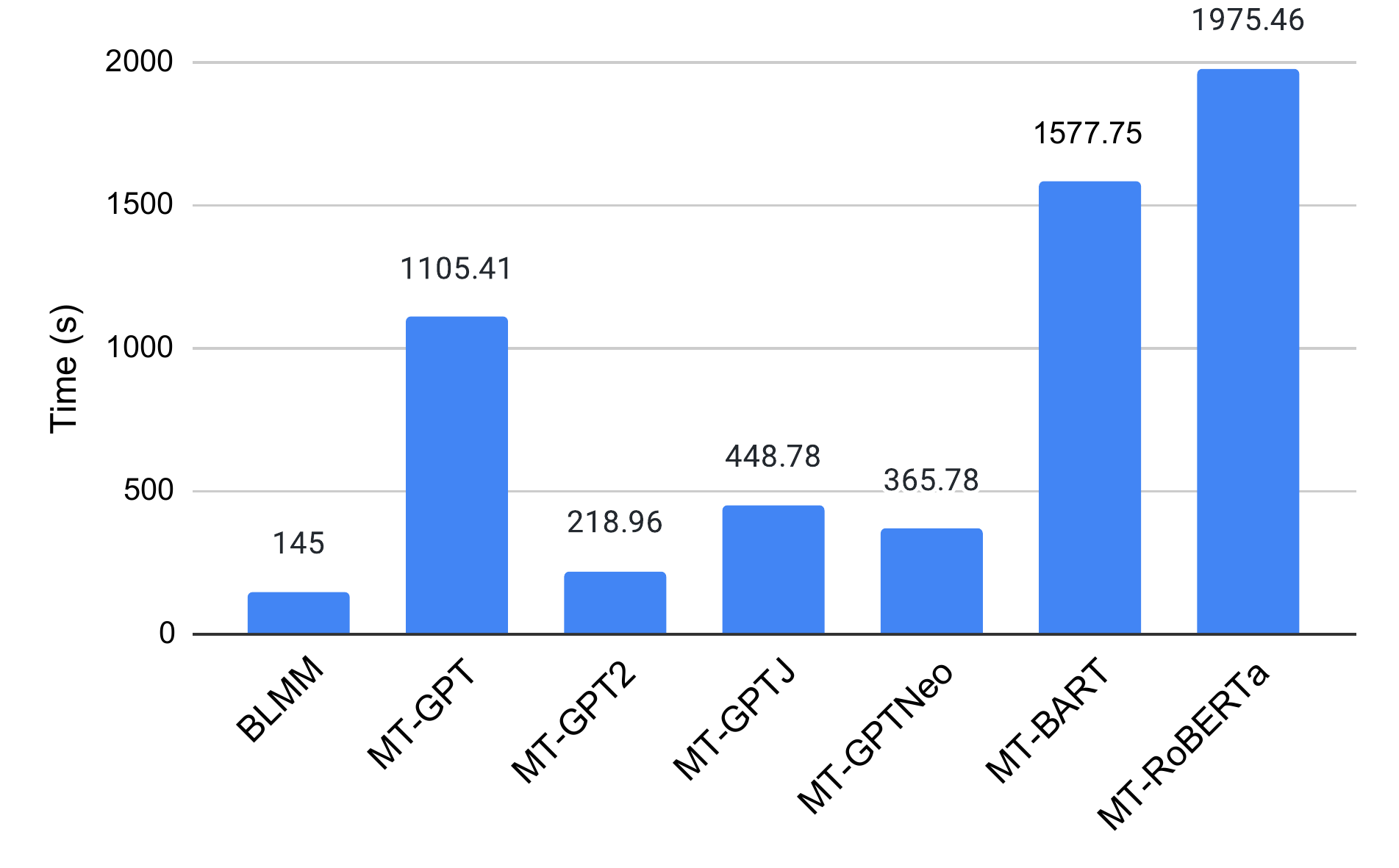}
  \caption{Time for generating 1,000 compositions by different models. The fastest model is BLMM while the slowest one is MT-RoBERTa.}
  \label{fig:time}
\end{figure}

We check the time complexity of the generators as shown in Figure \ref{fig:time}. We count the amount of time (in seconds) that each model needs to generate 1,000 compositions without any filtering. It is found that the BLMM model has the fastest generation with 145 seconds. The second fastest models are MT-GPT2, MT-GPTJ and MT-GPTNeo which use 219, 449, 366 seconds respectively. The slowest generators include MT-GPT, MT-BART, and MT-RoBERTa which are almost 7.6 to 13.6 times slower compared to BLMM. We also find that the generator models vary a lot in terms of generating qualified candidates that have <=30 atoms and <=8 elements. Figure S9 in the supplementary file shows the number of qualified candidate compositions generated by different models within 25,000 loops, each of which a sequence of 256 tokens is generated and then partitioned into multiple compositions. We find that MT-RoBERTa has an extremely low yield rate in generating qualified compositions compared to other transformer models. It is surprising that the MT-GPT model trained on the ICSD-pure dataset also has the difficulty to generate such qualified individuals. Both of them tend to generate formulas with >8 elements.

Another potential factor that affects the generator performance is the training set size. To check this issue, we train 15 MT-GPT2 models with training set sizes ranging from 1000 to 377,084. Their generation performances in terms of validity as represented by the CN and EB percentages are shown in the supplementary file Figure S7. A general trend we find is that increasing the training set size can lead to better generator performance. Another major decision in materials transformer training is to determine the optimal training epochs for each model over different datasets so that overfitting can be avoided. In the supplementary file, Figure S8 (b) shows the learning curves of training and validation errors for training MT-GPT2. It is found that after around 700 epochs, the training process starts to overfit as the validation loss begins to dominate the training loss. However, we observe that the corresponding CN/EB percentages do not degrade much despite the occurrence of the overfitting. 

\begin{figure}[ht!] 
    \begin{subfigure}[t]{0.5\textwidth}
        \includegraphics[width=1.0\textwidth]{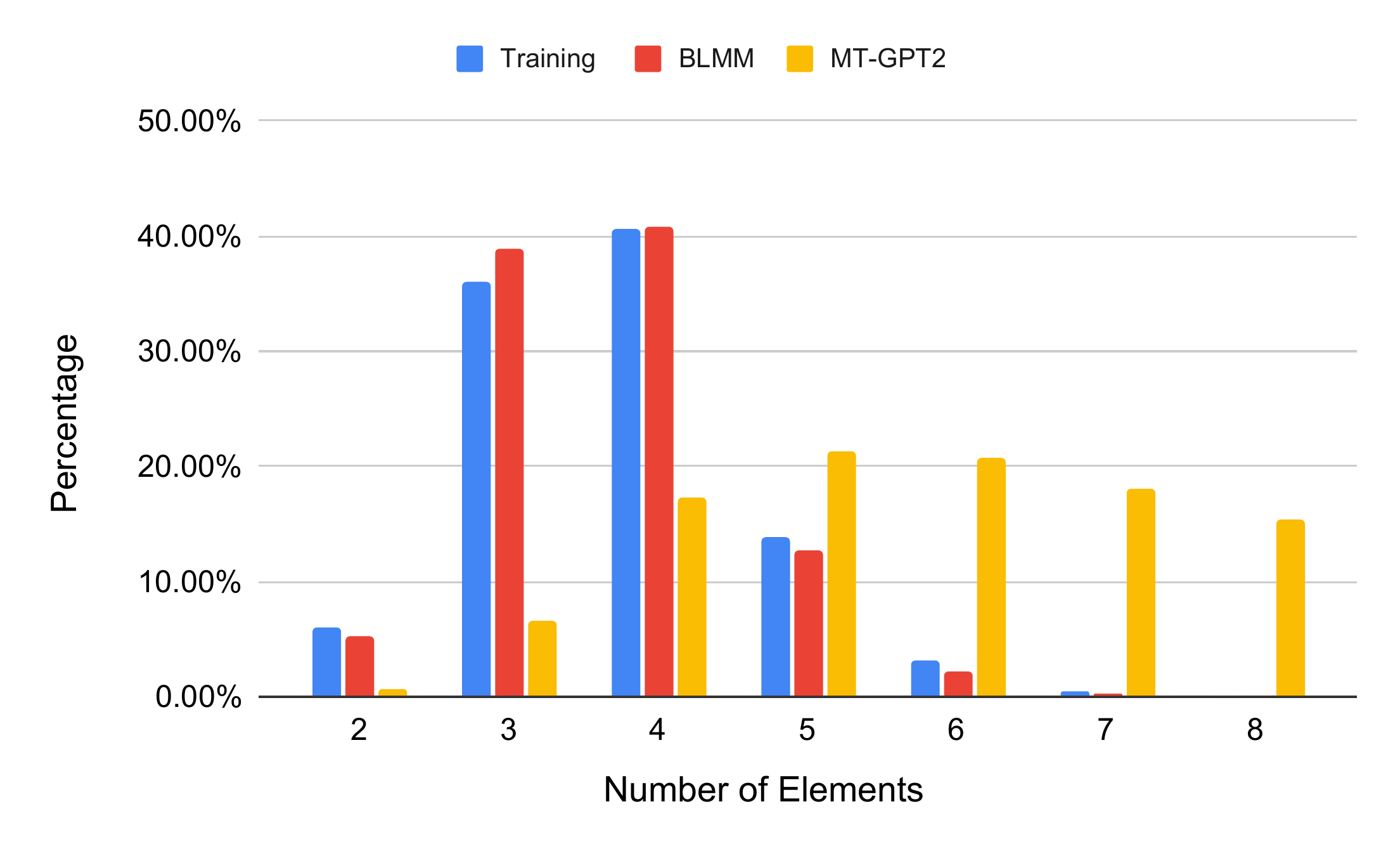}
        \caption{}
        \vspace{-3pt}
        \label{fig:elemenet_dist}
    \end{subfigure}
    \begin{subfigure}[t]{0.5\textwidth}
        \includegraphics[width=1.0\textwidth]{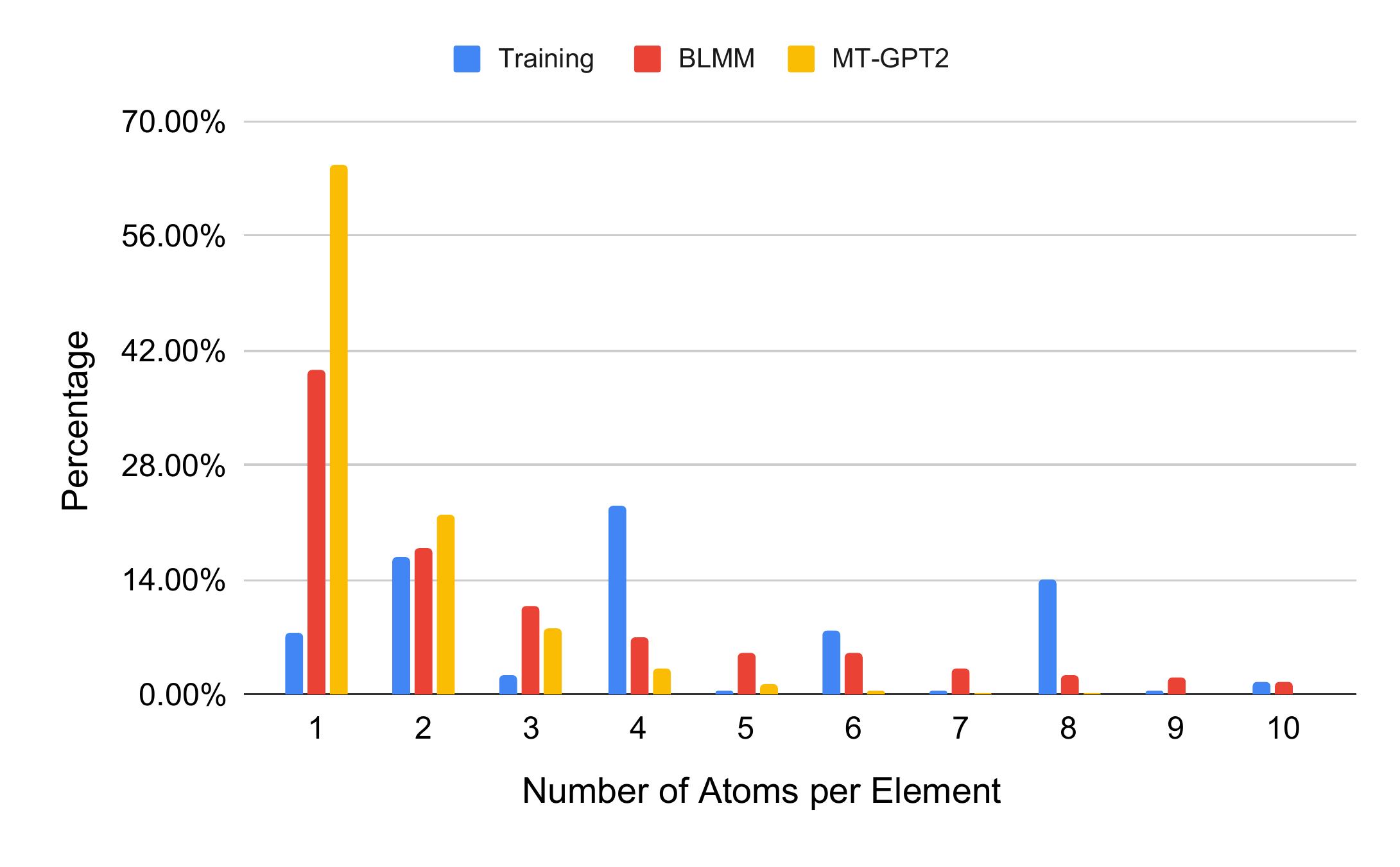}
        \caption{}
        \vspace{-3pt}
        \label{fig:atom_dist}
    \end{subfigure} 
    
     \begin{subfigure}[t]{1.0\textwidth}
        \includegraphics[width=1.0\textwidth]{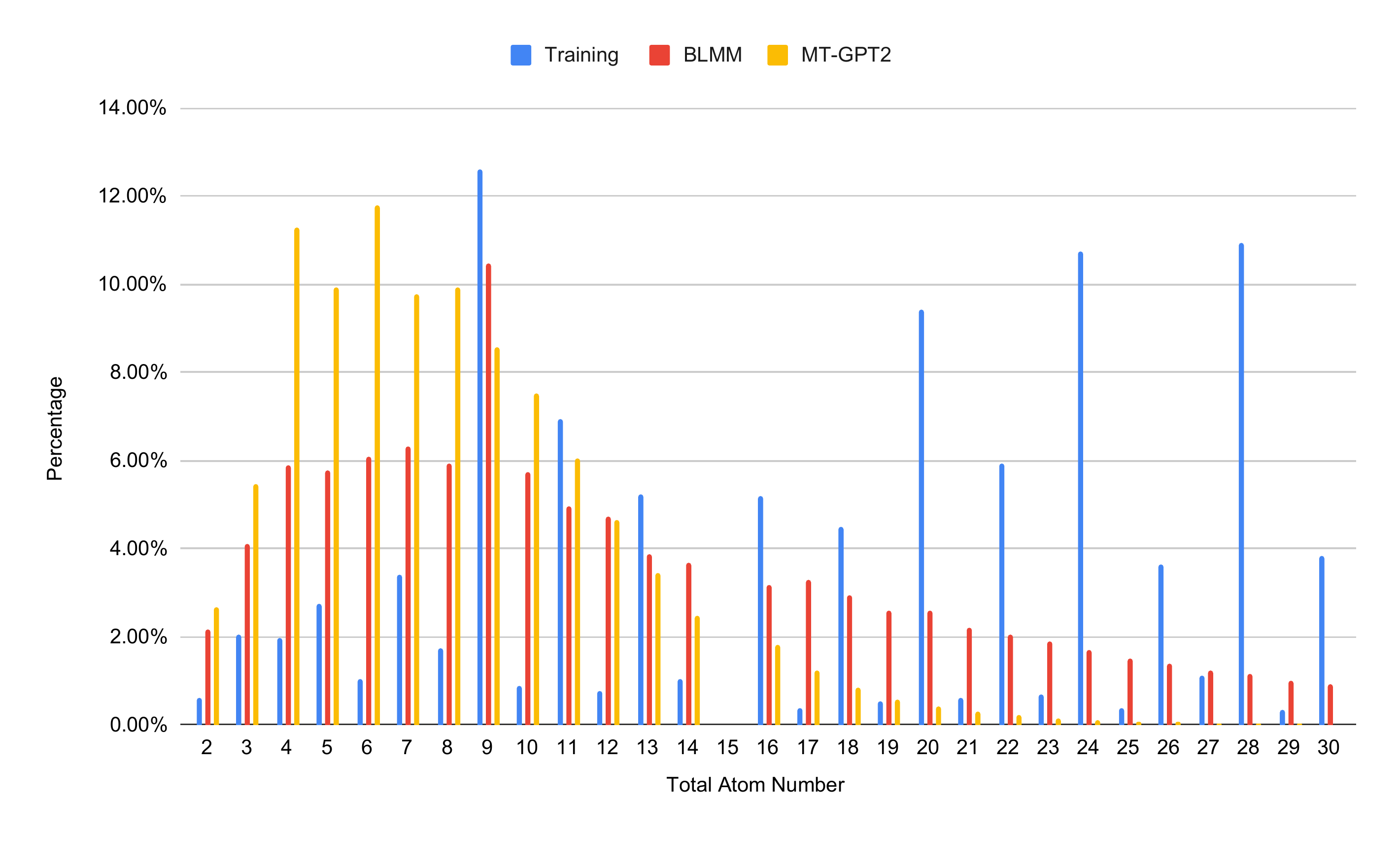}
        \caption{}
        \vspace{3pt}
        \label{fig:total_atom_dist}
    \end{subfigure} 
    
\caption{The comparison of the generation preferences of training samples, BLMM, and MT-GPT2 in terms of generated composition properties. (a) Distributions of the element numbers in each composition; (b) Distributions of the atom numbers for each element in all compositions. For the training set, the elements with more than 28 atoms are not counted in this plot. (c) Distribution of total atom number for each composition. For the training set, the compositions with more than 30 atoms are not plotted in this figure.}
  \label{fig:elment_atom_dist}
\end{figure}

To further examine the generation capabilities and potential bias of different materials transformers, we train the BLMM and MT-GPT2 models using the Bandgap-30K dataset as defined in Section \ref{subsec:bandgap} (Conditional generation) and  generate 89,321 and 87,232 samples respectively. We then plot the distributions of the number of elements within compositions, the number of atoms per element, and the total number of atoms within a composition as shown in Figure \ref{fig:elment_atom_dist}. First, we find that the distribution of the number of elements within the training compositions are highly imbalanced (Figure \ref{fig:elemenet_dist}) with the highest percentage of ternary and quaternary materials followed by quinary and binary materials. There are very few samples with more than 6 elements. It is then interesting to find that the element number distribution of the BLMM-generated samples follows closely to the training set, indicating that the BLMM model tends to learn well of the chemical patterns from the training set. On the other hand, the MT-GPT2-generated samples have a very different distribution in terms of the element number of compositions: it tends to generate a large proportion of samples with more than four elements. 

Next, Figure \ref{fig:atom_dist} shows the distribution of the number of atoms per element within the samples. The training samples have a variety of atom numbers per element ranging from 1 to more than 10, with relatively small percentages of ones. In contrast, the samples generated by BLMM contain much more compositions with 1-atom elements. However, the MT-GPT2 model is even more biased as it tends to generate most of single-atom elements in the compositions (65\%) followed by 2-atom elements (16\%) (check yellow bars). It has a low probability to generate samples with more than four-atom elements. We find the generation preferences of MT-GPT2 also apply to all other transformer-based models except BLMM. 

We further check the distribution of the total atoms within a composition as shown in Figure \ref{fig:total_atom_dist}. First, it is found that the training set has many samples with a large number of atoms with peaks at 9, 20, 24, and 28. In contrast, the total atom numbers of the samples generated by GPT2 peak at 4, 5, 6, 7, 8, and 9, which are much smaller than the training set. It can barely generate samples with total atom numbers more than 20. On the other hand, the BLMM model has a more balanced generation capability: while it also peaks at 9, 4, 5, 6, 7, 8, 10, and 11 in terms of the total atom number within compositions, it can still generate large percentages of samples with more than 20 atoms. It is also interesting to notice that there are no materials with 15 atoms within the formula.

\FloatBarrier

Our extensive experiments show that the transformer-based language models can generate chemically valid hypothetical material compositions as shown by the high percentages of generated charge neutral and balanced electronegativity samples. However, it is not clear whether these compositions can be synthesized into stable structures. This especially may become a concern when a generative model tends to generate compositions with a large number of elements. While several machine learning models have been developed for synthesizability prediction \cite{jang2020structure}, formation energy prediction \cite{omee2022scalable}, and e-above-hull calculation, these models and algorithms usually require the availability of the crystal structures which are not available for composition generators that we propose here. To do that, we can use the recently developed template-based crystal structure prediction algorithms \cite{kusaba2022crystal,wei2021tcsp} or the deep learning-based \cite{hu2021alphacrystal}, and global optimization-based crystal structure prediction tools \cite{oganov2012crystal,shao2022symmetry} to predict the crystal structures for the generated hypothetical compositions by our materials transformer models. Together, we are able to explore and discover new materials in a much larger area of the almost infinite chemical design space.

\section{Materials and Methods}
\label{sec:others}

\subsection{Dataset}
\label{subsec:dataset}

To evaluate the performance of our language model-based generators, we prepare six different datasets: Hybrid-mix, Hybrid-pure, Hybrid-strict, ICSD-mix, ICSD-pure, and Bandgap-30K. 

The formulas in the Hybrid-mix dataset are selected from the ICSD/MP/OQMD databases with the number of elements less than 9, the number of atoms in the unit cell less than 100, and without fractional atom numbers for any element in the formula. While the Hybrid-mix dataset may contain a certain amount of materials that are not charge neutral or balanced electronegativity, the Hybrid-pure dataset has samples selected from the Hybrid-mix dataset, which are charge neutral and have balanced electronegativity. The Hybrid-strict dataset is obtained using a similar method to the Hybrid-pure dataset, but we use the strict ICSD oxidation states of the elements to calculate the CN and EB, which is different (with more strict constraints) from the Hybrid-pure dataset and the dataset used in our previous study \cite{wei2022crystal}. 

For the two ICSD datasets, the formulas in the ICSD-mix dataset are sampled from the ICSD database with the number of elements less than 9, the number of atoms in unit cell less than 100, and without fractional coordinates. It may contain formulas that do not meet the CN and EB criteria. Samples in the ICSD-pure dataset are selected from the ICSD-mix dataset, which satisfy CN and EB rules. 

In addition, for the experiment of band gap prediction, we prepare a Bandgap-30K dataset, which contains 30,000 formulas from the MP database with band gaps above 1.98 eV. For those formulas with multiple phases, we include it if it has one phase with the band gap greater than 2.0 eV. 

Overall, we get the Hybrid-mix, Hybrid-pure, Hybrid-strict, ICSD-mix, ICSD-pure, and Bandgap-30K dataset with total of 418983, 257138, 212778, 52317, 39431, and 30000 samples respectively. We divide all datasets into the training set, the test set, and the validation set in a ratio of 9/0.5/0.5, and the detailed number of each set is shown in Table \ref{tab:datasets}.

\begin{table}[]
\centering
\caption{Six datasets used in experiments. For Hybrid and ICSD datasets, -pure datasets only include selected samples with the neutral charge and balanced electronegativity; -mix datasets do not have such limits.}
\label{tab:datasets}
\vspace{4pt}
\begin{tabular}{|c|ccc|cc|c|}
\hline
\rowcolor[HTML]{FFCC67} 
\cellcolor[HTML]{FFFFFF}                   & \multicolumn{3}{c|}{\cellcolor[HTML]{FFCC67}Hybrid datasets (ICSD+MP+OQMD)}        & \multicolumn{2}{c|}{\cellcolor[HTML]{FFCC67}ICSD datasets} & MP dataset  \\ \cline{2-7} 
\multirow{-2}{*}{\cellcolor[HTML]{FFFFFF}} & \multicolumn{1}{c|}{Hybrid-mix} & \multicolumn{1}{c|}{Hybrid-pure} & Hybrid-strict & \multicolumn{1}{c|}{ICSD-mix}          & ICSD-pure         & Bandgap-30K \\ \hline
Total                                      & \multicolumn{1}{c|}{418983}     & \multicolumn{1}{c|}{257138}      & 212778        & \multicolumn{1}{c|}{52317}             & 39431             & 30000       \\ \hline
Train                                      & \multicolumn{1}{c|}{398033}     & \multicolumn{1}{c|}{244281}      & 202139        & \multicolumn{1}{c|}{50755}             & 37459             & 28500       \\ \hline
Valid                                      & \multicolumn{1}{c|}{10475}      & \multicolumn{1}{c|}{6428}        & 5319          & \multicolumn{1}{c|}{1336}              & 986               & 750         \\ \hline
Test                                       & \multicolumn{1}{c|}{10475}      & \multicolumn{1}{c|}{6429}        & 5320          & \multicolumn{1}{c|}{1336}              & 986               & 750         \\ \hline
\end{tabular}
\end{table}

\subsection{Pseudo-random composition generator}
We build a pseudo-random composition generator as the baseline generation model. For all generated samples, we count the numbers of samples with different numbers of elements from 2 to 8. Then for each sample with the number of $K$ elements, we generate the same number of composition samples with $K$ elements. For these samples with $K$ elements, We randomly pick the atom number from 1 to 20 for each of these $K$ elements. This process ensures the distribution of binary, ternary, etc.

\subsection{DFT calculations}
\label{subsec:DFT}

We use the first-principles calculations based on the density functional theory (DFT) using the Vienna \textit{ab initio} simulation package (VASP) to check the structural stability of the predicted materials \cite{Vasp1,Vasp2,Vasp3,Vasp4}. The projected augmented wave (PAW) pseudo-potentials, where 520 eV plane-wave cutoff energy, are used to treat the electron-ion interactions \cite{PAW1, PAW2}. The exchange-correlation functional is considered with the generalized gradient approximation (GGA) based on the Perdew-Burke-Ernzerhof (PBE) method \cite{GGA1, GGA2}. The energy convergence criterion is set as 10$^{-5}$ eV, while the atomic positions are optimized with the force convergence criterion of 10$^{-2}$ eV/{\AA}. The Brillouin zone integration for the unit cells was computed using the $\Gamma$-centered  Monkhorst-Pack $k$-meshes. The Formation energies (in eV/atom) of several materials are determined based on the expression in  Eq.~\ref{eq:form}, where $E[\mathrm{Material}]$ is the total energy per unit formula of the considered structure, $E[\textrm{A}_i]$ is the energy of $i^\mathrm{th}$ element of the material, $x_i$ indicates the number of A$_i$ atoms in a unit formula, and $n$ is the total number of atoms in a unit formula($n=\sum_i x_i$).

\begin{equation}
    E_{\mathrm{form}} =\frac{1}{n}(E[\mathrm{Material}] - \sum_i x_i E[\textrm{A}_i])
    \label{eq:form}
\end{equation}

\subsection{Evaluation criteria}
\label{subsec:criteria}

There are three main criteria we used in this paper, the validity, the uniqueness, the recovery rate and the novelty \cite{dan2020generative}.

\textbf{Validity.} First, the two basic chemical rules of crystals, charge neutrality (CN) and balanced electronegativity (EB) percentages, are used to evaluate the validity of generated formulas. We use the method to calculate CN and EB proposed in \cite{davies2019smact} to obtain the percentages of generated samples that conform to these two rules. In addition, we check the stability of the generated samples to evaluate their validity performance using their predicted formation energies. The higher-energy region they are in, the lower quality they have.

\textbf{Uniqueness.} The uniqueness represents the ability of a generative model to generate unique formulas, which is used to calculate the percentage of the number of unique samples in the whole generated samples. The higher the uniqueness, the more diverse samples the model can produce.

\textbf{Recovery Rate and Novelty.} The recovery rate is used to estimate the percentage of generated formulas from the training set or the test set. The samples in the training set and test set are known, which means that the high recovery rate shows that this generative model has high performance on discovery materials. Another criterion related to the recovery rate is the novelty. The novelty of a generative model measures the percentage of the generated formulas that do not exist in the training or test sets.

\subsection{Hyper-parameters Tuning}
\label{subsec:para-tuning}

Since the content and quantity of Hybrid datasets and ICSD datasets are very different, we tuned hyper-parameters for the Hybrid-mix dataset and the ICSD-pure dataset to choose appropriate hyper-parameters for Hybrid datasets and ICSD datasets respectively. Using the Hybrid-mix and the ICSD-pure datasets, we evaluated how the key hyper-parameters affect the generation performance on the MT-GPT2, the MT-BART, and the MT-RoBERTa models. We trained these models with different hyper-parameters, then generated about 10,000 formulas to evaluate their performance using two criteria, CN (charge neutral) and EB (balanced electronegativity). To find suitable parameters, We set default parameters for each model, then changed one of them each time to train models and evaluated their performance. In this section, we only go to the detail of the hyper-parameter tuning of the MT-GPT2 model on the Hybrid-mix dataset (More details on the hyper-parameter tuning of other models are based on the supplementary file Figure S1-S5). For the MT-GPT2 model, the hyper-parameters we set include the maximum length of the formula tokens that this model might ever be used with, the dimensionality of the embeddings and hidden states, the number of hidden layers in the transformer encoder, and the number of attention heads for each attention layer in the transformer encoder.

\begin{figure}[hb!] 
    \begin{subfigure}[t]{0.50\textwidth}
        \includegraphics[width=\textwidth]{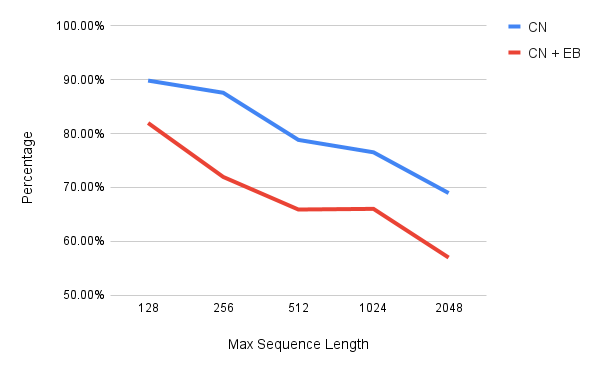}
        \caption{}
        \vspace{-3pt}
        \label{fig:gpt2_hy_po}
    \end{subfigure}   
    \begin{subfigure}[t]{0.50\textwidth}
        \includegraphics[width=\textwidth]{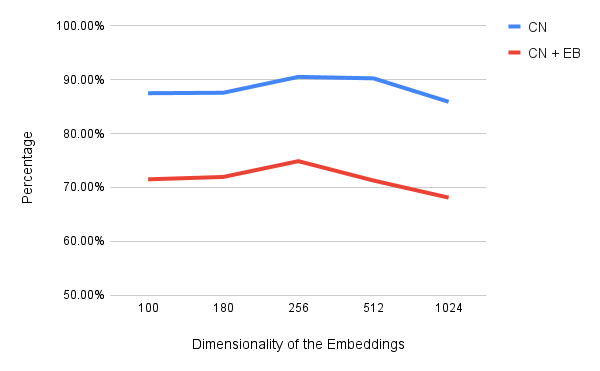}
        \caption{}
        \vspace{-3pt}
        \label{fig:gpt2_hy_emb}
    \end{subfigure}
    \begin{subfigure}[t]{0.50\textwidth}
        \includegraphics[width=\textwidth]{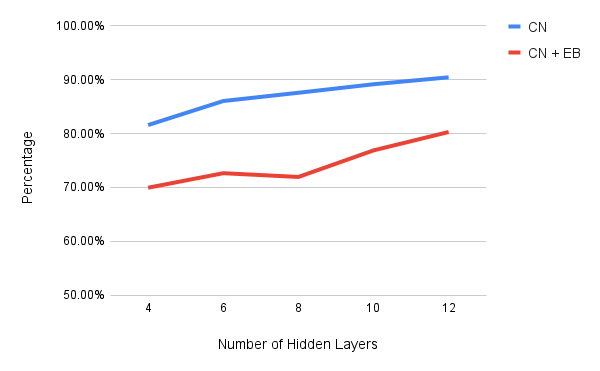}
        \caption{}
        \vspace{-3pt}
        \label{fig:gpt2_hy_layer}
    \end{subfigure}
    \begin{subfigure}[t]{0.50\textwidth}
        \includegraphics[width=\textwidth]{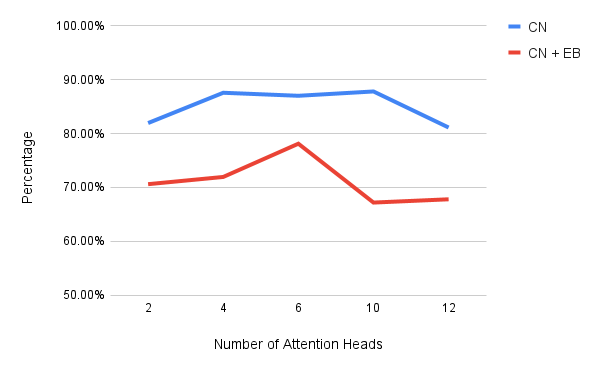}
        \caption{}
        \vspace{-3pt}
        \label{fig:gpt2_hy_head}
    \end{subfigure}  
   \caption{Hyper-parameter tuning of MT-GPT2 model on the Hybrid-mix dataset. (a)-(d) show the percentages of charge-neutral (CN in the figures) and balanced electronegativity (CN+EB in the figures) out of all generated samples by the MT-GPT2 models trained on the Hybrid-mix dataset with different maximum sequence lengths, different dimensionalities of the embeddings, different numbers of the hidden layers in the Transformer encoder, and different number of attention heads for each attention layer in the Transformer encoder.}
  \label{fig:hyperparameter}
\end{figure}

\textbf{The maximum length of the formula tokens}. The default value we set is 256. As shown in Figure \ref{fig:gpt2_hy_po}, we evaluated different lengths, 128, 512, 1024, and 2048. Taking both CN and EB percentages into account, models on the Hybrid-mix dataset can hit a better performance with the max length of 128, which achieves the CN percentage of 89.81\% and the EB percentage of 81.95\%. As the max length increases, the CN and EB percentages of generated formulas gradually decrease.

\textbf{Embeddings dimension.} The default value of the embedding dimension is 180. As shown in Figure \ref{fig:gpt2_hy_emb}, different values were set as 100, 256, 512, and 1024. Then models on the Hybrid-mix dataset achieved the CN percentages of 87.47\%, 87.55\%, 90.49\%, 90.22\%, 85.88\% and EB percentages of 71.51\%, 71.95\%, 74.88\%, 71.29\%, 68.12\% for varying embedding dimension from small to large respectively. When the embedding dimension is 256, the CN and EB percentages can reach 90.49\% and 74.88\%, respectively.

\textbf{The number of hidden layers.} We set the default value as 8, and evaluated models with 4, 6, 10, and 12 hidden layers. As shown in \ref{fig:gpt2_hy_layer}, as the number of hidden layers increases, the model can learn more about the formulas in the training set, then can generate formulas with higher CN and EB percentages. Therefore, the best CN and EN percentages can be 90.41\% and 80.28\% when the number of hidden layers is 12. 

\textbf{The number of attention heads.} We increase the number of attention heads from 2 to 12, and set the default value as 4. It is observed from Figure \ref{fig:gpt2_hy_head} that the best result is achieved with 6 heads. It is worth mentioning that when we set the number of attention heads, we need to make sure that the embedding dimension is a multiple of the number of attention heads.

For each model, we perform similar tuning procedures as described above, and the final hyper-parameters of each model are shown in Table S1 from the supplementary file.

 In addition, We wonder if different training sizes might affect the performance of generated formulas. Therefore, in addition to the tuning of hyper-parameters for models, we also conducted experiments to evaluate the effect of different sizes of the training set. We changed the size of the Hybrid-mix dataset from 1,000 to 377,083 (90\% of the Hybrid-mix dataset) to train different MT-GPT2 models, then used the trained models to generate 10,000 formulas. Figure S6 in the supplementary file shows that although the curve is a bit bumpy, the proportion of eligible formulas in the generated formulas shows an overall upward trend.

\subsection{Formation energy and band gap prediction models based on Roost}
\label{subsec:formation_energy}

To evaluate the generator performances, we train two composition-based machine learning models. Both formation energy and band gap prediction use the dataset downloaded from the MP database \cite{jain2013commentary}. The machine learning model we used is based on Roost, a graph message passing neural network as described in \cite{goodall2020predicting}. The training set of Roost-FE (formation energy) contains 125,613 unique compositions. For those compositions with multiple phases, we only keep the records with the lowest formation energies. The Roost-Bandgap model is trained with 113,501 samples. The formation energy roost model achieves an MAE of 70.181 eV while the band gap predictor achieves an MAE of 0.6645 eV as evaluated on the 10\% hold-out test sets.

\FloatBarrier

\bibliographystyle{unsrt}  
\bibliography{references}  

\section*{Acknowledgement}

\paragraph{Funding:}The research reported in this work was supported in part by National Science Foundation under the grant and 1940099 and 1905775. The views, perspectives, and content do not necessarily represent the official views of the NSF. 
\paragraph{Author contributions:}
Conceptualization, J.H.; methodology,J.H., N.F., L.W., Q.L, D.S., Y.S.; software, N.F., J.H., L.W., Y.S., Q.L.; resources, J.H.; writing--original draft preparation, J.H., F.N., L.W., D.S., R.X., S.S.O.; writing--review and editing, J.H, F.N., L.W., D.S.; visualization, N.F., J.H., L.W., Y.S., Q.L, D.S.; supervision, J.H.;  funding acquisition, J.H.
\paragraph{Competing interests: The authors declare that they have no competing interests}

\end{document}